# The application of radiation diffuse scattering to the calculation of phase diagrams of F.C.C. substitutional alloys

V.A. Tatarenko*, T.M. Radchenko

*Institute for Metal Physics, N.A.S. of the Ukraine, 36 Academician Vernadsky Blvd., UA-03680 Kyiv-142, Ukraine*

**Abstract**

By using quantitative information about the radiation diffuse-scattering intensity of the disordered f.c.c. substitutional alloy $Me'_{1\text{-}c}Me''_c$ ($c$—concentration) the Fourier component, $\tilde{w}(\mathbf{k})$, of mixing energies of $Me'$ and $Me''$ atoms may be estimated. We have to use the measurement data of the diffuse-scattering intensities at the corresponding reciprocal-space points $\mathbf{k}$ of the disordered phase and then determine the parameter $\tilde{w}(\mathbf{k})$. The statistical thermodynamics of the non-ideal solid solution is determined by these energy parameters $\{\tilde{w}(\mathbf{k})\}$. Therefore, one can obtain the configuration free energy of an alloy, $F=U\text{-}TS$ ($U$—internal energy, $S$—entropy), and then determine its fundamental thermodynamic characteristics, including not only its phase diagram, but also the concentration-dependent order–disorder transformation temperature, temperature and concentration long-range order parameter dependences, chemical activity, heat capacity etc. Some thermodynamic properties are calculated within the framework of the statistical-thermodynamic approach for f.c.c.-Ni–Fe alloy. The diffuse-scattering intensity values are taken from data in the literature.

## 1. Introduction

The method to consider the interatomic long-range interactions in the statistical theory of solid solutions was proposed by Khachaturyan [1–4]. Even in the long-range interactions all the thermodynamical functions can be evaluated in the self-consistent field approximation if we know just a few energetic parameters—$\tilde{w}(\mathbf{k})$, which are the Fourier transforms of the interchange energies ('mixing' energies), $w(\mathbf{r})$, taken in the non-equivalent points [4–7];

$$\tilde{w}(\mathbf{k}) = \sum_{\mathbf{r}} w(\mathbf{r})\, \mathrm{e}^{-i\mathbf{k}\cdot\mathbf{r}}, \quad (1)$$

here $\mathbf{r}=\mathbf{R}\text{–}\mathbf{R}'$ ($\mathbf{R}$, $\mathbf{R}'$—crystal-lattice sites vectors), the reciprocal space vectors $\mathbf{k}$ are related to the positions of fundamental ($\mathbf{k}=\mathbf{0}$) and superstructure ($\mathbf{k}=\mathbf{k}_1,\mathbf{k}_2,\ldots,\mathbf{k}_s,\ldots,\mathbf{k}_M$) reciprocal-lattice sites (in the first Brillouin zone) of the 'star', $s$, of wave vectors (M—number of sublattices formed during the ordering). The sum in Eq. (1) is taken over all vector differences $\mathbf{R}\text{–}\mathbf{R}'$ of crystal lattice.

The number of the energy parameters $\tilde{w}(\mathbf{k})$ is determined by the structure of the ordered phase. For instance, in the self-consistent field approximation, the statistical thermodynamics of an f.c.c. Ni–Fe alloy [variation of $\tilde{w}(\mathbf{rk})$ with a number of co-ordination shell $l$, at different annealing temperatures for this alloy, is represented in Fig. 1] are completely determined by two energy parameters: $\tilde{w}(\mathbf{k}_1)$ and $\tilde{w}(\mathbf{0})$, where $\mathbf{k}_1=2\pi\mathbf{a}_3^*$ and $\mathbf{k}_2=\mathbf{0}$ are the wave vectors corresponding to the (001) and (000) reciprocal-lattice points of such an alloy, respectively; $\mathbf{a}_3^*$ is the basis vector of the reciprocal lattice in the [001] direction. Thus, all the interchange energies in the arbitrary-range interaction model are contained in two energy parameters $\tilde{w}(\mathbf{k}_1)$ and $\tilde{w}(\mathbf{0})$.

Quantitative information about the radiation diffuse-scattering intensity, $T_{\mathrm{diff}}(\mathbf{q})$, measured at a point $\mathbf{q}$, which is disposed at a distance $\mathbf{k}$ from $2\pi\mathbf{B}$—nearest reciprocal-space site ($\mathbf{B}$—reciprocal-lattice vector) of the disordered f.c.c. substitutional alloy $Me'_{1\text{-}c}Me''_c$ ($c$—concentration) at the absolute temperature $T$, may be used to estimate the Fourier component, $\tilde{w}(\mathbf{k})$, of 'mixing' energies of $Me'$ and $Me''$ atoms [4,5]:

$$I_{\text{diff}}(\mathbf{q}) \propto D\frac{c(1-c)}{1+c(1-c)\tilde{w}(\mathbf{k})/(k_BT)}, \tag{2}$$

where $D$ is normalizing factor and $k_B$ is the Boltzmanm constant. Thus, we have an unique possibility of direct experimental determination of energy parameters $\tilde{w}(\mathbf{k})$ of a system. We have to use the measurement data of the diffuse-scattering intensities at the corresponding reciprocal-space points $\mathbf{k}$ of the disordered phase and then determine the parameter $\tilde{w}(\mathbf{k})$ by using Eq. (2). As was mentioned above, the statistical thermodynamics of the non-ideal solid solution is determined by these energy parameters $\{\tilde{w}(\mathbf{k})\}$.

* Corresponding author Supported by STCU#2412.
*E-mail address:* tatar@imp.kiev.ua (V.A. Tatarenko).

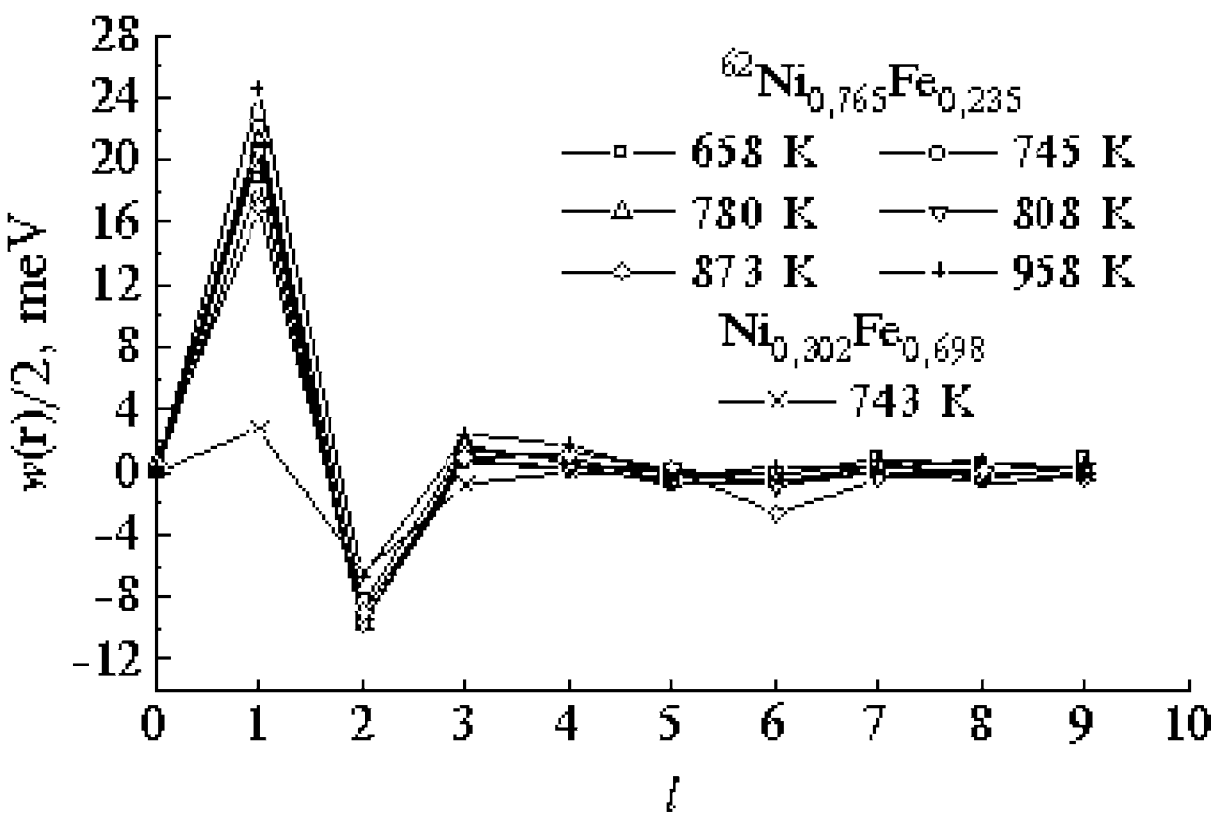

Fig. 1. Variation of 'mixing' energy $w(\mathbf{r})$ with number of co-ordination shell $l$ at different annealing temperatures for f.c.c-Ni–Fe alloy (within the approximation $w(\mathbf{r})|_{r=0}=0$ and without taking into account atomic size mismatch effects) [7].

## 2. Model

In the self-consistent field approximation, the configuration free energy of the binary alloy is [4,5]

$$F_{\text{conf}} = U_{\text{conf}} - TS_{\text{conf}}, \tag{3}$$

where

$$U_{\text{conf}} = \frac{1}{2}\sum_{\mathbf{R}}\sum_{\mathbf{R}'} w(\mathbf{R}-\mathbf{R}')P(\mathbf{R})P(\mathbf{R}') \tag{4}$$

is the configuration internal energy,

$$S_{\text{conf}} = -k_B\sum_{\mathbf{R}}\{P(\mathbf{R})\ln P(\mathbf{R})[1-P(\mathbf{R})]\ \ln[1-P(\mathbf{R})]\} \tag{5}$$

is the configuration entropy. The function $P(\mathbf{R})$ ($P(\mathbf{R}')$) in Eqs. (4) and (5) is the occupation probability of finding an atom of the definite kind (for example a Fe atom) on the site $\mathbf{R}$ ($\mathbf{R}'$) of the f.c.c. lattice.

Using the Khachaturyan method we can represent the occupation probabilities for Fe atoms of superstructure $L1_2$-type in the form of a superposition of several static concentration waves:

$$P(\mathbf{R}) = c_{\text{Fe}} + \frac{\eta}{4} \times \left[\exp(i2\pi\mathbf{a}_1^*\cdot\mathbf{R}) + \exp(i2\pi\mathbf{a}_2^*\cdot\mathbf{R}) + \exp(i2\pi\mathbf{a}_3^*\cdot\mathbf{R})\right], \tag{6}$$

where η is the long-range order parameter and $\mathbf{a}_1^*$, $\mathbf{a}_2^*$, $\mathbf{a}_3^*$ are the basis vectors of the reciprocal lattice in [100], [010], [001] directions.

By substitution of Eq. (6) into Eqs. (4) and (5) one can obtain an expression for the configuration internal energy,

$$U_{\text{conf}} = \frac{N}{2}\left\{\tilde{w}(\mathbf{0})c_{\text{Fe}}^2 + \frac{3}{16}\eta^2\tilde{w}(\mathbf{k}_1)\right\}, \tag{7}$$

where $N$ is the total number of the f.c.c. lattice sites, and the configuration entropy is

$$\begin{aligned} S_{\text{conf}}(\eta) = -\frac{k_BN}{4}\Big\{&3\left(c_{\text{Fe}}-\frac{\eta}{4}\right)\ln\left(c_{\text{Fe}}-\frac{\eta}{4}\right) \\ &+3\left(1-c_{\text{Fe}}+\frac{\eta}{4}\right)\ln\left(1-c_{\text{Fe}}+\frac{\eta}{4}\right) \\ &+\left(c_{\text{Fe}}+\frac{3}{4}\eta\right)\ln\left(c_{\text{Fe}}+\frac{3}{4}\eta\right) \\ &+\left(1-c_{\text{Fe}}-\frac{3}{4}\eta\right)\ln\left(1-c_{\text{Fe}}-\frac{3}{4}\eta\right)\Big\}. \end{aligned} \tag{8}$$

Eqs. (3), (7) and (8) lead to the following:

$$\begin{aligned} \frac{F_{\text{conf}}}{N} = &\frac{1}{2}\left\{\tilde{w}(\mathbf{0})c_{\text{Fe}}^2 + \frac{3}{16}\eta^2\tilde{w}(\mathbf{k}_1)\right\} \\ &+\frac{k_BT}{4}\Big\{3\left(c_{\text{Fe}}-\frac{\eta}{4}\right)\ln\left(c_{\text{Fe}}-\frac{\eta}{4}\right) \\ &+3\left(1-c_{\text{Fe}}+\frac{\eta}{4}\right)\ln\left(1-c_{\text{Fe}}+\frac{\eta}{4}\right) \\ &+\left(c_{\text{Fe}}+\frac{3}{4}\eta\right)\ln\left(c_{\text{Fe}}+\frac{3}{4}\eta\right) \\ &+\left(1-c_{\text{Fe}}-\frac{3}{4}\eta\right)\ln\left(1-c_{\text{Fe}}-\frac{3}{4}\eta\right)\Big\}. \end{aligned} \tag{9}$$

The presence of magnetism in Ni–Fe alloys essentially complicates the analysis of this system. Expression for the total 'mixing' energy, $w(\mathbf{r})=w(\mathbf{R}-\mathbf{R})$, is

$$w(\mathbf{R}-\mathbf{R}') = w_{prm}(\mathbf{R}-\mathbf{R}') + J_{FeFe}(\mathbf{R}-\mathbf{R}')s_{Fe}^2\sigma_{Fe}^2 + J_{NiNi}(\mathbf{R}-\mathbf{R}')s_{Ni}^2\sigma_{Ni}^2 - 2J_{FeNi}(\mathbf{R}-\mathbf{R}')s_{Fe}s_{Ni}\sigma_{Fe}\sigma_{Ni} \quad (10a)$$

in the real space, or

$$\tilde{w}(\mathbf{k}) = \tilde{w}_{prm}(\mathbf{k}) + \tilde{J}_{FeFe}(\mathbf{k})s_{Fe}^2\sigma_{Fe}^2 + \tilde{J}_{NiNi}(\mathbf{k})s_{Ni}^2\sigma_{Ni}^2 - 2\tilde{J}_{FeNi}(\mathbf{k})s_{Fe}s_{Ni}\sigma_{Fe}\sigma_{Ni} \quad (10b)$$

in the reciprocal space, where $w_{prm}(\mathbf{R}-\mathbf{R}')$ is the 'mixing' energy in 'paramagnetic' condition, $J_{\alpha\beta}(\mathbf{R}-\mathbf{R})$ is the exchange energy ($\alpha,\beta$ = Fe,Ni), $s_{Fe}$ and $s_{Ni}$ are the total spins of Fe and Ni atoms respectively, $\sigma_{Fe}$ and $\sigma_{Ni}$ are the relative magnetizations of Fe and Ni atoms respectively, $\tilde{w}_{prm}(\mathbf{k})$ is the Fourier-component of 'mixing' energy in 'paramagnetic' condition, and $J_{\alpha\beta}(\mathbf{k}) = \sum_{\mathbf{r}} J_{\alpha\beta}(\mathbf{r})\, e^{-i\mathbf{k}\cdot\mathbf{r}}$ is the Fourier-component of exchange energy.

So the *total* 'configuration-dependent' free energy $F = F_{W\mathrm{conf}+F\mathrm{magn}} = U - TS$ ($F_{magn}$—magnetic free energy, $U$—total 'configuration-dependent' internal energy, $S$—total 'configuration-dependent' entropy) for $Ni_3Fe$ alloy is as follows (see also [4, 8–11] and their references):

$$\begin{aligned}\frac{F}{N} = {} & \frac{1}{2}\Big[\tilde{w}_{prm}(\mathbf{0})(1-c_{Ni})^2 + \tilde{J}_{FeFe}(\mathbf{0})c_{Fe}^2\sigma_{Fe}^2 s_{Fe}^2 \\ & + \tilde{J}_{NiNi}(\mathbf{0})(1-c_{Fe})^2\sigma_{Ni}^2 s_{Ni}^2 \\ & + 2\tilde{J}_{FeNi}(\mathbf{0})c_{Fe}(1-c_{Fe})\sigma_{Fe}s_{Fe}\sigma_{Ni}s_{Ni} + \frac{3}{16\eta}{}^2 \\ & \Big(\tilde{w}_{prm}(\mathbf{k}_1) + \tilde{J}_{FeFe}(\mathbf{k}_1)\sigma_{Fe}^2 s_{Fe}^2 \\ & + \tilde{J}_{NiNi}(\mathbf{k}_1)\sigma_{Ni}^2 s_{Ni}^2 - 2\tilde{J}_{FeNi}(\mathbf{k}_1)\sigma_{Fe}\sigma_{Ni}s_{Fe}s_{Ni}\Big)\Big] \\ & + \frac{k_B T}{4}\Big[3\Big(1-c_{Fe}+\frac{\eta}{4}\Big)\ln\Big(1-c_{Fe}+\frac{\eta}{4}\Big) \\ & + 3\Big(c_{Fe}-\frac{\eta}{4}\Big)\ln\Big(c_{Fe}-\frac{\eta}{4}\Big) \\ & + \Big(1-c_{Fe}-\frac{3}{4}\eta\Big)\ln\Big(1-c_{Fe}-\frac{3}{4}\eta\Big) \\ & + \Big(c_{Fe}+\frac{3}{4}\eta\Big)\ln\Big(c_{Fe}+\frac{3}{4}\eta\Big)\Big] \\ & - k_B T(1-c_{Fe})\Big[\ln\ \mathrm{sh}\Big(\Big(1+\frac{1}{2s_{Ni}}\Big)y_{Ni}(\sigma_{Ni})\Big) \\ & - \ln\ \mathrm{sh}\Big(\frac{1}{2s_{Ni}}y_{Ni}(\sigma_{Ni})\Big) - \sigma_{Ni}y_{Ni}(\sigma_{Ni})\Big] \\ & - k_B T c_{Fe}\ln\ \mathrm{sh}\Big(1+\frac{1}{2s_{Fe}}\Big)y_{Fe}(\sigma_{Fe})\Big) \\ & - \ln\ \mathrm{sh}\Big(\frac{1}{2s_{Fe}}y_{Fe}(\sigma_{Fe})\Big) - \sigma_{Fe}y_{Fe}(\sigma_{Fe})\Big],\end{aligned} \quad (11)$$

here $y_\alpha = \frac{s_\alpha H_{eff}}{k_B T}$, where $H_{eff} = -g\mu_B\Gamma\sigma_\alpha$ ($g$—Lande factor, $\mu_B$—Bohr magneton, $\Gamma$—coefficient of Weis 'molecular' field).

Eq. (11) and conditions $\partial F/\partial\eta = 0$, $\partial F/\partial\sigma_{Fe} = 0$, $\partial F/\partial\sigma_{Ni} = 0$ lead to the following transcendental equations:

$$\begin{aligned}\ln & \frac{\Big(c_{Fe}-\frac{\eta}{4}\Big)\Big(1-c_{Fe}-\frac{3}{4}\eta\Big)}{\Big(c_{Fe}+\frac{3}{4}\eta\Big)\Big(1-c_{Fe}+\frac{\eta}{4}\Big)} \\ & = \frac{\eta}{k_B T}\Big[\tilde{w}_{prm}(\mathbf{k}_1) + \tilde{J}_{FeFe}(\mathbf{k}_1)\sigma_{Fe}^2 s_{Fe}^2 \\ & + \tilde{J}_{NiNi}(\mathbf{k}_1)\sigma_{Ni}^2 s_{Ni}^2 \\ & - 2\tilde{J}_{FeNi}(\mathbf{k}_1)\sigma_{Fe}\sigma_{Ni}s_{Fe}s_{Ni}\Big], \\ \sigma_{Ni} & = \mathrm{B}_{s_{Ni}}\Big(-\frac{1}{(1-c_{Fe})k_B T}\Big\{\tilde{J}_{NiNi}(\mathbf{0}) \\ & (1-c_{Fe})^2 s_{Ni}^2\sigma_{Ni} + \\ & \tilde{J}_{FeNi}(\mathbf{0})c_{Fe}(1-c_{Fe})s_{Fe}s_{Ni}\sigma_{Fe} \\ & + \frac{3}{16}\eta^2\Big[\tilde{J}_{NiNi}(\mathbf{k}_1)s_{Ni}^2\sigma_{Ni} - \tilde{J}_{FeNi}(\mathbf{k}_1)s_{Fe}s_{Ni}\sigma_{Fe}\Big]\Big\}\Big) \\ \sigma_{Fe} & = \mathrm{B}_{s_{Fe}}\Big(-\frac{1}{c_{Fe}k_B T}\Big(\tilde{J}_{FeFe}\Big(\mathbf{0}\Big)c_{W\mathrm{fFe}}s_{Fe}^2\sigma_{Fe} \\ & + \tilde{J}_{FeNi}(\mathbf{0})c_{Fe}(1-c_{Fe})s_{Fe}s_{Ni}\sigma_{Ni} \\ & + \frac{3}{16}\eta^2\Big[\tilde{J}_{FeFe}(\mathbf{k}_1)s_{Fe}^2\sigma_{Fe} - \tilde{J}_{FeNi}(\mathbf{k}_1)s_{Fe}s_{Ni}\sigma_{Ni}\Big]\Big)\Big),\end{aligned} \quad (12)$$

here $\hat{\mathrm{B}}_{s\alpha}$ is the Brillouin function, which is equal [9,10] to

$$\mathrm{B}_{s_\alpha}(y_\alpha) = \Big(1+\frac{1}{2s_\alpha}\Big)\mathrm{cth}\Big[\Big(1+\frac{1}{2s_\alpha}\Big)y_\alpha\Big] - \frac{1}{2s_\alpha}\mathrm{cth}\Big[\frac{1}{2s_\alpha}y_\alpha\Big]. \quad (13)$$

Relative magnetization may be defined as

$$\sigma_\alpha = \mathrm{B}_{s_\alpha}\Big(\frac{s_\alpha H_{eff}}{k_B T}\Big). \quad (14)$$

Eq. (12) lets us calculate the long-range order parameter $\eta$ and relative magnetizations $\sigma_{Fe}$ and $\sigma_{Ni}$, corresponding to the temperature $T$ and composition $c$ (for example, concentration $c_{Fe}$). Substituting these parameters, $\eta$, $\sigma_{Fe}$ and $\sigma_{Ni}$, into Eq. (11), we can obtain the dependence of the free energy $F$ on temperature and concentration and consequently can construct the equilibrium diagram.

Previously mentioned Eq. (2) may be used to estimate the Fourier components, $\tilde{w}(\mathbf{k})$, of 'mixing' energies. However, it is necessary to mean that energy parameters $\tilde{w}(\mathbf{k}_1)$ and $\tilde{w}(\mathbf{0})$ are not constants in a ferromagnetic state. Except temperature-dependent (implicitly) contribution $\tilde{w}_{prm}(\mathbf{k})$, they contain additive components dependent on temperature and composition of alloy by way of squares of magnetisation. To exclude these additive components, we have to measure the intensity of the diffuse scattering of radiation (X-ray or thermal neutrons) only in the paramagnetic region of the equilibrium diagram (at temperature $T > T_C$, where $T_C$ is the Curie temperature).

## 3. Calculation of 'mixing' energy, order–disorder transition temperature and critical parameters of alloy

In a mean field approximation, the value $T/I_{\text{diff}}(001)$ is a linear in $T$; so the expression for diffuse scattering intensity is [12]

$$\frac{T}{I_{\text{diff}}(001)} = \frac{T}{C_0} + \frac{c(1-c)\tilde{w}(001)}{C_0 k_B}$$

where the normalization constant $C_0$ is calculated in order to ensure that [12]

$$\int \tilde{w}(\mathbf{k})\mathrm{l}\mathbf{k} = 0 \text{ and } C_0 = \left| \int \frac{\mathrm{d}\mathbf{k}}{I_{\text{diff}}(\mathbf{k})} \right|^{-1}$$

$$(2\pi)^3/\mathbf{a}_1^* \cdot [\mathbf{a}_2^* \times \mathbf{a}_3^*]/ \text{ or } \mathrm{D} = 1[5, 7, 8].$$

Value of $\tilde{w}(\mathbf{k}_1)/k_B$ for f.c.c.-$^{62}Ni_{0.765}Fe_{0.235}$ was easily estimated by the using experimental results of [12]. It is $\tilde{w}(\mathbf{k}_1)/k_B = -4218$ K, that is $\tilde{w}(\mathbf{k}_1) = -0.86$ eV.

The disorder to order phase transition for $Ni_3Fe$ alloy is transition of the 1st order. It means that Kurnakov temperature $T_K$ can be obtained from Eq. (12) and the following condition [4]: $N^{-1}\Delta F(\Delta\eta, T_K) = 0$ or

$$\begin{aligned}\frac{\Delta F(\Delta\eta, T_K)}{N} = \frac{1}{N}\left\{F(\Delta\eta, T_K) - F(0, T_K)\right\} = \frac{1}{2}\frac{3}{16}\Delta\eta^2\tilde{w}(\mathbf{k}_1) + \\ + \frac{k_B T_K}{4}\left(3w_f c_{\text{Fe}} - \frac{\Delta\eta}{4}\right)\ln\left(c_{\text{Fe}} - \frac{\Delta\eta}{4}\right) \\ + 3\left(1 - c_{\text{Fe}} + \frac{\Delta\eta}{4}\right)\ln\left(1 - c_{\text{Fe}} + \frac{\Delta\eta}{4}\right) \\ + \left(c_{\text{Fe}} + \frac{3}{4}\Delta\eta\right)\ln\left(c_{\text{Fe}} + \frac{3}{4}\Delta\eta\right) \\ + \left(1 - c_{\text{Fe}} - \frac{3}{4}\Delta\eta\right)\ln\left(1 - c_{\text{Fe}} - \frac{3}{4}\Delta\eta\right)w_f \\ - k_B T_K\left(c_{\text{Fe}}\ln c_{\text{Fe}} + (1 - c_{\text{Fe}})\ln(1 - c_{\text{Fe}})\right).\end{aligned} \quad (15)$$

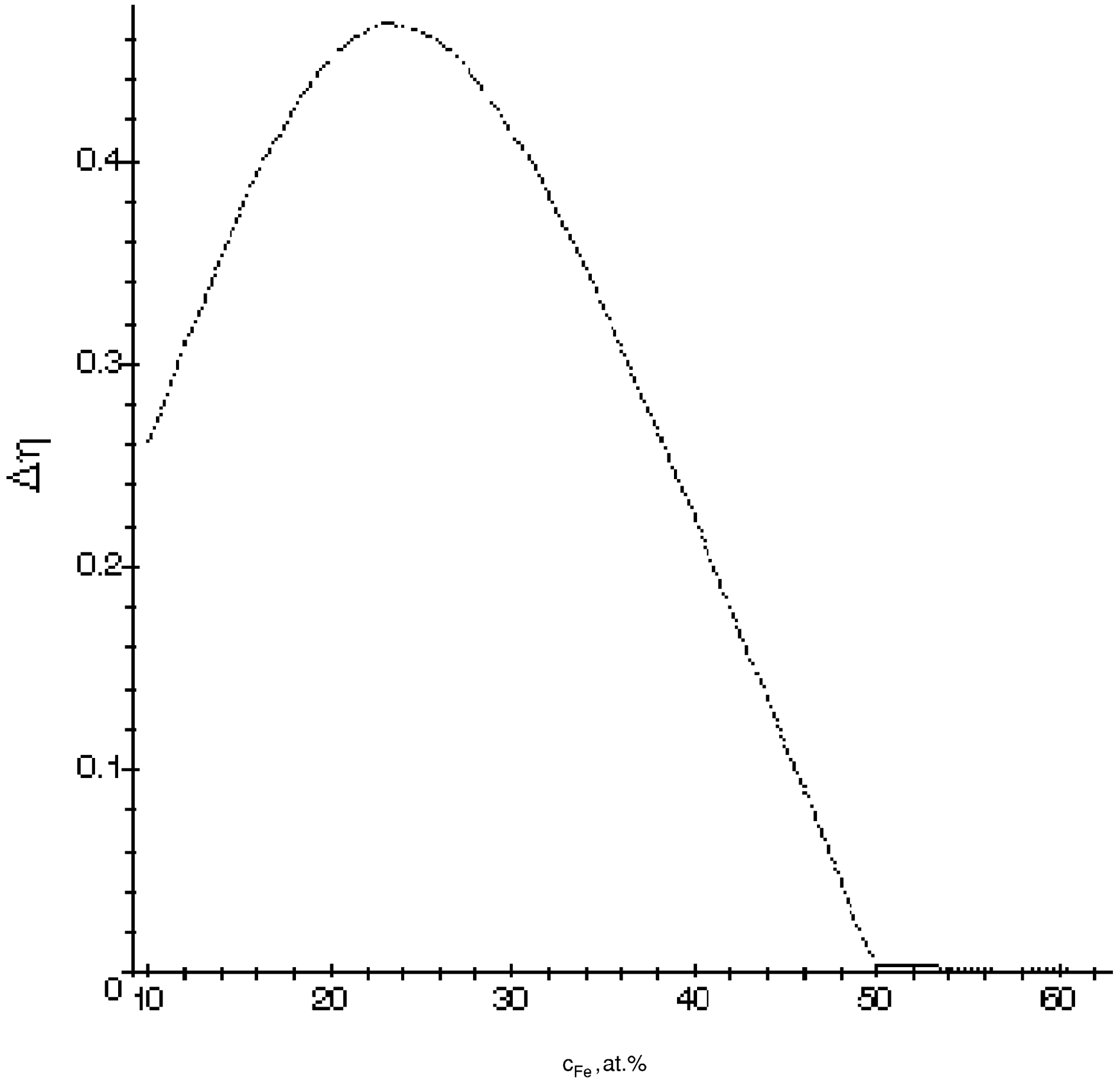

Fig. 2. Calculated theoretical dependence of long-range order parameter jump, $\Delta\eta$, of $L1_2$-type superstructure on Fe content, $c_{\text{Fe}}$, at corresponding Kurnakov temperatures for f.c.c.-Ni–Fe alloy.

Thus, to obtain the jump of the long-range order parameter of $L1_2$-type superstructure we have to use numerical computing to solve the equation

$$\frac{N}{2}\frac{3}{16}\Delta\eta\ln\frac{\left(c_{\mathrm{Fe}}-\frac{\Delta\eta}{4}\right)\left(1-c_{\mathrm{Fe}}-\frac{3}{4}\Delta\eta\right)}{\left(c_{\mathrm{Ni}}+\frac{3}{4}\Delta\eta\right)\left(1-c_{\mathrm{Fe}}+\frac{\Delta\eta}{4}\right)} + \frac{N}{4}\left(3\left(c_{\mathrm{Fe}}-\frac{\Delta\eta}{4}\right)\ln\left(c_{\mathrm{Fe}}-\frac{\Delta\eta}{4}\right) + 3\left(1-c_{\mathrm{Fe}}+\frac{\Delta\eta}{4}\right)\ln\left(1-c_{\mathrm{Fe}}+\frac{\Delta\eta}{4}\right) + \left(c_{\mathrm{Fe}}+\frac{3}{4}\Delta\eta\right)\ln\left(c_{\mathrm{Fe}}+\frac{3}{4}\Delta\eta\right) + \left(1-c_{\mathrm{Fe}}-\frac{3}{4}\Delta\eta\right)\ln\left(1-c_{\mathrm{Fe}}-\frac{3}{4}\Delta\eta\right)\right) - N(c_{\mathrm{Fe}}\ln c_{\mathrm{Fe}}+(1-c_{\mathrm{Fe}})\ln(1-c_{\mathrm{Fe}})) = 0 \quad (16)$$

and to substitute its solution (that is dependence $\Delta\eta(c_{\mathrm{Fe}})$) to the following expression for the Kurnakov temperature obtaining:

$$\tau_K = \frac{\Delta\eta}{\ln\frac{\left(c_{\mathrm{Fe}}-\frac{\Delta\eta}{4}\right)\left(1-c_{\mathrm{Fe}}-\frac{3}{4}\Delta\eta\right)}{\left(c_{\mathrm{Fe}}+\frac{3}{4}\Delta\eta\right)\left(1-c_{\mathrm{Fe}}+\frac{\Delta\eta}{4}\right)}} \quad (17)$$

where $\boldsymbol{\tau_K} = \frac{k_B T_K}{\bar{w}(\mathbf{k}_1)}$.

Dependencies $\Delta\eta(c_{\mathrm{Fe}})$ and $\tau_K(c_{\mathrm{Fe}})$ are represented on Figs. 2 and 3. The comparatively low value of the jump of the long-range order parameter at $T_k$ (viz $\Delta\eta(c_{\mathrm{Fe}})\leqslant 0{,}47$) for the $Ni_3Fe$ superstructure is evidence of the stability of this ordered phase in respect to the short-range ordered phase and in some way justifies the neglect of the correlation effects in the spatial locations of atoms of components (in the vicinity of Kurnakov temperature at least).

To obtain concentration Kurnakov (absolute) temperature dependence we have to know the concen-tration dependence of magnetizations (for example from Eq. (12)) at the same Kurnakov temperature. That is to say, it is necessary to also solve two transcendental combined equations relative to magnetizations, $s_{\mathrm{Ni}}$ and $s_{\mathrm{Fe}}$, at $\dot{O}_{\hat{E}}(c_{\mathrm{Ni}})$:

$$\sigma_{\mathbf{Ni}} = \mathrm{B}_{s_{\mathrm{Ni}}} \times \Bigg( -\frac{1}{(1-c_{\mathbf{Fe}})k_B\Big[\tilde{w}_{\mathrm{prm}}(\mathbf{k}) + J_{\mathbf{FeFe}}(\mathbf{k})\sigma_{\mathbf{Fe}}^2 s_{\mathbf{Fe}}^2 + \tilde{J}_{\mathbf{NiNi}}(\mathbf{k})\sigma_{\mathbf{Ni}}^2 s_{\mathbf{Ni}}^2 - 2\tilde{J}_{\mathbf{FeNi}}(\mathbf{k})\sigma_{\mathbf{Fe}}\sigma_{\mathbf{Ni}} s_{\mathbf{Fe}} s_{\mathbf{Ni}}\Big]\tau_K(c_{\mathbf{Ni}})} \times \Big\{\tilde{J}_{\mathbf{NiNi}}(\mathbf{0})(1-c_{\mathbf{Fe}})^2 s_{\mathbf{N}i}^2\sigma_{\mathbf{Ni}} + \tilde{J}_{\mathbf{FeNi}}(\mathbf{0})c_{\mathbf{Fe}}(1-c_{\mathbf{Fe}})s_{\mathbf{Fe}}s_{\mathbf{Ni}}\sigma_{\mathbf{Fe}} + \frac{3}{16}\eta_K^2(c_{\mathbf{Ni}})\Big[\tilde{J}_{\mathbf{NiNi}}(\mathbf{k})s_{\mathbf{Ni}}^2\sigma_{\mathbf{Ni}} - \tilde{J}_{\mathbf{FeNi}}(\mathbf{k})s_{\mathbf{Fe}}s_{\mathbf{Ni}}\sigma_{\mathbf{Fe}}\Big]\Big\}\Bigg) \tag{18a}$$

$$\sigma_{\mathbf{Fe}} = \mathrm{B}_{s_{\mathrm{Fe}}} \times \Bigg( -\frac{1}{c_{\mathbf{Fe}}k_B\Big[\tilde{w}_{\mathrm{prm}}(\mathbf{k}) + \tilde{J}_{\mathbf{FeFe}}(\mathbf{k})\sigma_{\mathbf{Fe}}^2 s_{\mathbf{Fe}}^2 + \tilde{J}_{\mathbf{NiNi}}(\mathbf{k})\sigma_{\mathbf{Ni}}^2 s_{\mathbf{Ni}}^2 - 2\tilde{J}_{\mathbf{FeNi}}(\mathbf{k})\sigma_{\mathbf{Fe}}\sigma_{\mathbf{Ni}} s_{\mathbf{Fe}} s_{\mathbf{Ni}}\Big]\tau_K(c_{\mathbf{Ni}})} \times \Big(\tilde{J}_{\mathbf{FeFe}}(\mathbf{0})c_{\mathbf{Fe}} s_{\mathbf{Fe}}^2\sigma_{\mathbf{Fe}} + \tilde{J}_{\mathbf{FeNi}}\Big(\mathbf{0}\Big)c_{\mathbf{Fe}}(1-c_{\mathbf{Fe}})s_{\mathbf{Fe}}s_{\mathbf{Ni}}\sigma_{\mathbf{Ni}} + \frac{3}{16}\eta_K^2(c_{\mathbf{Fe}})\Big[\tilde{J}_{\mathbf{FeFe}}(\mathbf{k})s_{\mathbf{Fe}}^2\sigma_{\mathbf{Fe}} - \tilde{J}_{\mathbf{FeNi}}(\mathbf{k})s_{\mathbf{Fe}}s_{\mathbf{Ni}}\sigma_{\mathbf{Ni}}\Big]\Big)\Bigg). \tag{18b}$$

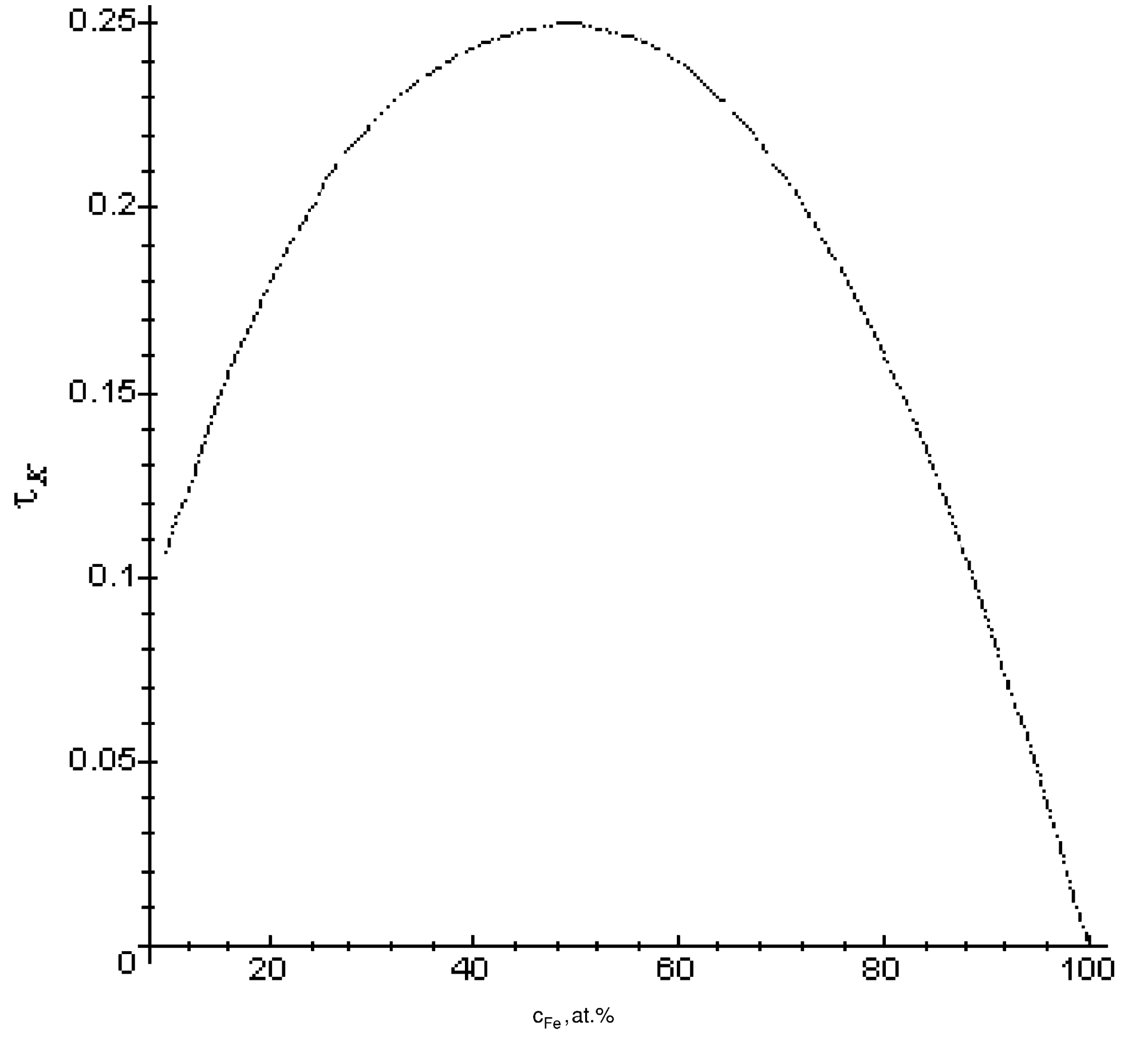

Fig. 3. Calculated theoretical dependence of module of reduced Kurnakov temperature, $\tau_K$, of superstructure $L1_2$-type on Fe content, $c_{\mathrm{Fe}}$, for f.c.c.-Ni–Fe alloy.

From Eq. (18a,b) (within the framework of the approximation of $\tilde{J}_{\alpha\beta}(\mathbf{k}) \approx -\frac{1}{3}\tilde{J}_{\alpha\alpha'}(\mathbf{0})$ (α, β = Fe, Ni), which is correct for the Fourier-components corresponding to k points (100)-type of the f.c.c.-lattice reciprocal space, as well as within the framework of the assumption of $s_{\mathrm{Fe}} = 3$ and $s_{\mathrm{Ni}} = 1$) we can calculate the dependencies $\sigma_{\mathrm{Fe}}(c_{\mathrm{Fe}})$, $\sigma_{\mathrm{Ni}}(c_{\mathrm{Fe}})$ and then construct graphs to determine an absolute value of the Kurnakov temperature

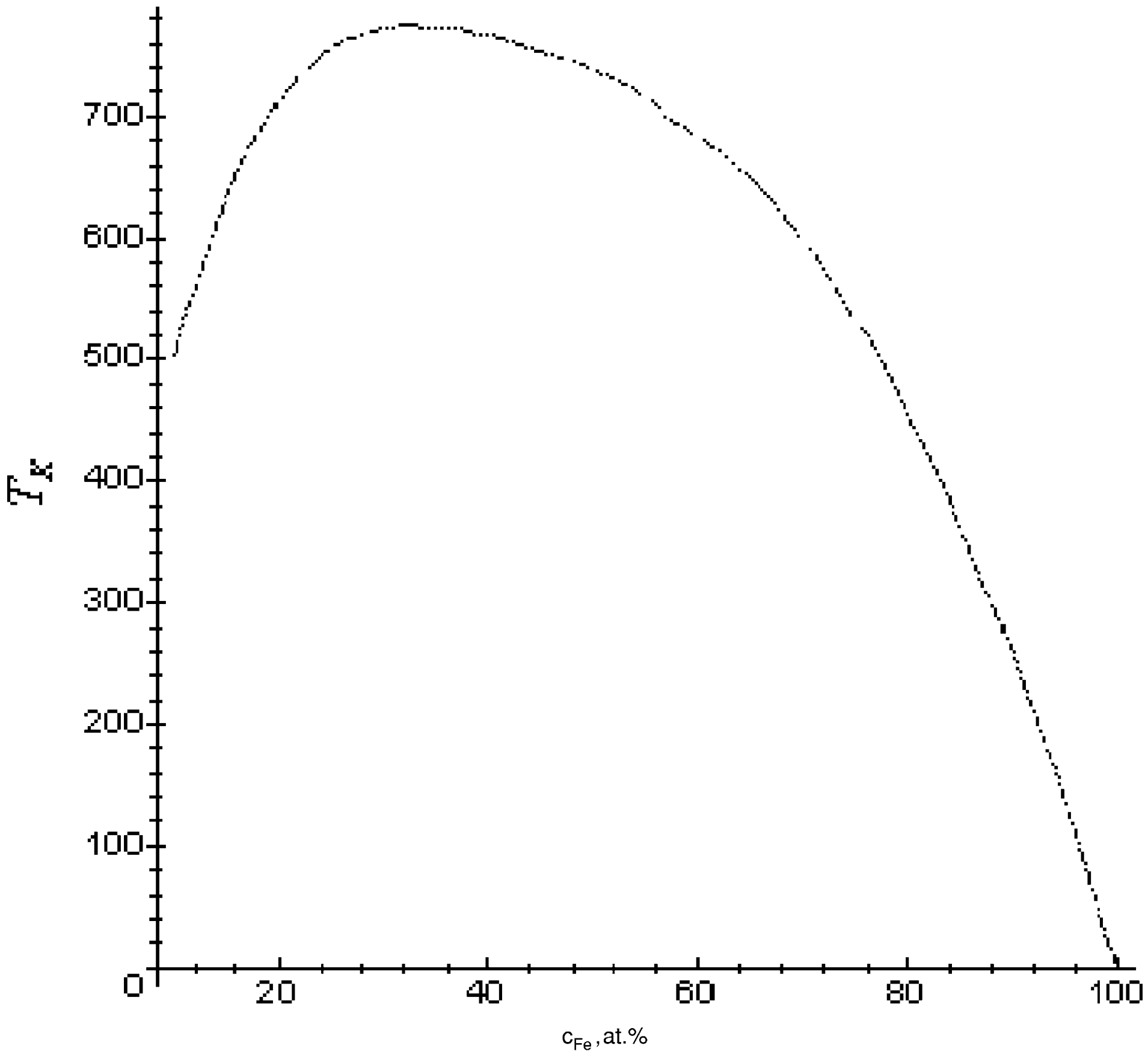

Fig. 4. Calculated theoretical dependence of Kurnakov temperature, $T_K$, of superstructure $L1_2$-type on Fe content, $c_{\mathrm{Fe}}$, for f.c.c.-Ni–Fe alloy (taking into consideration influence of long-range ferromagnetic order).

$$T_K(c_{\rm Fe}) = \Big\{\tilde{w}_{\rm prm}(\mathbf{k}_1) + \tilde{J}_{\rm FeFe}(\mathbf{k}_1)s_{\rm Fe}^2\sigma_{\rm Fe}^2 + \tilde{J}_{\rm NiNi}(\mathbf{k}_1)s_{\rm Ni}^2\sigma_{\rm Ni}^2 - 2\tilde{J}_{\rm FeNi}(\mathbf{k}_1)s_{\rm Fe}s_{\rm Ni}\sigma_{\rm Fe}\sigma_{\rm Ni}\Big\}\tau_K(c_{\rm Fe})/\mathbf{k}_\beta \qquad (19)$$

and an average 'atomic' magnetic moment of alloy at this Kurnakov temperature,

$$\overline{\mu} = c_{\rm Fe}s_{\rm Fe}\sigma_{\rm Fe} + (1 - c_{\rm Fe})s_{\rm Ni}\sigma_{\rm Ni}, \qquad (20)$$

which are represented on Figs. 4 and 5.

In the end, one can compare, for example, the graph on Fig. 4 with corresponding experimental data relative to Kurnakov temperatures for superstructure of the $L1_2$-type, with numerical and experimental data for corresponding $\tau_K$ (see [13–18] and Fig. 6), and also with analytical results of preliminary asymptotic high-temperature expansion relative to $\tau_K(c_{\rm Fe})$ [6]. The coherence is satisfactory.

## 4. Conclusions

The method of radiation diffuse scattering by single crystals of disordered alloys can be successfully applied for construction of equilibrium diagrams, for determining dependences of long-range order parameters on temperature and concentration, for calculating chemical activities, heat capacity, temperatures of the order–disorder transition etc. The advantages of the method are as follows:

- (i) the opportunity of taking into account the arbitrary-range interatomic interactions and not only the nearest and next-nearest neighbour interactions;
- (ii) the consistent account of the magnetism to free energy within the framework of the self-consistent field approximation;
- (iii) it is not necessary to use fitting parameters, which are usually used to tie together evaluated and observed results (see, for example, [7,19,20]);
- (iv) within the framework of the present approach one can easily take into account the correlation effects.

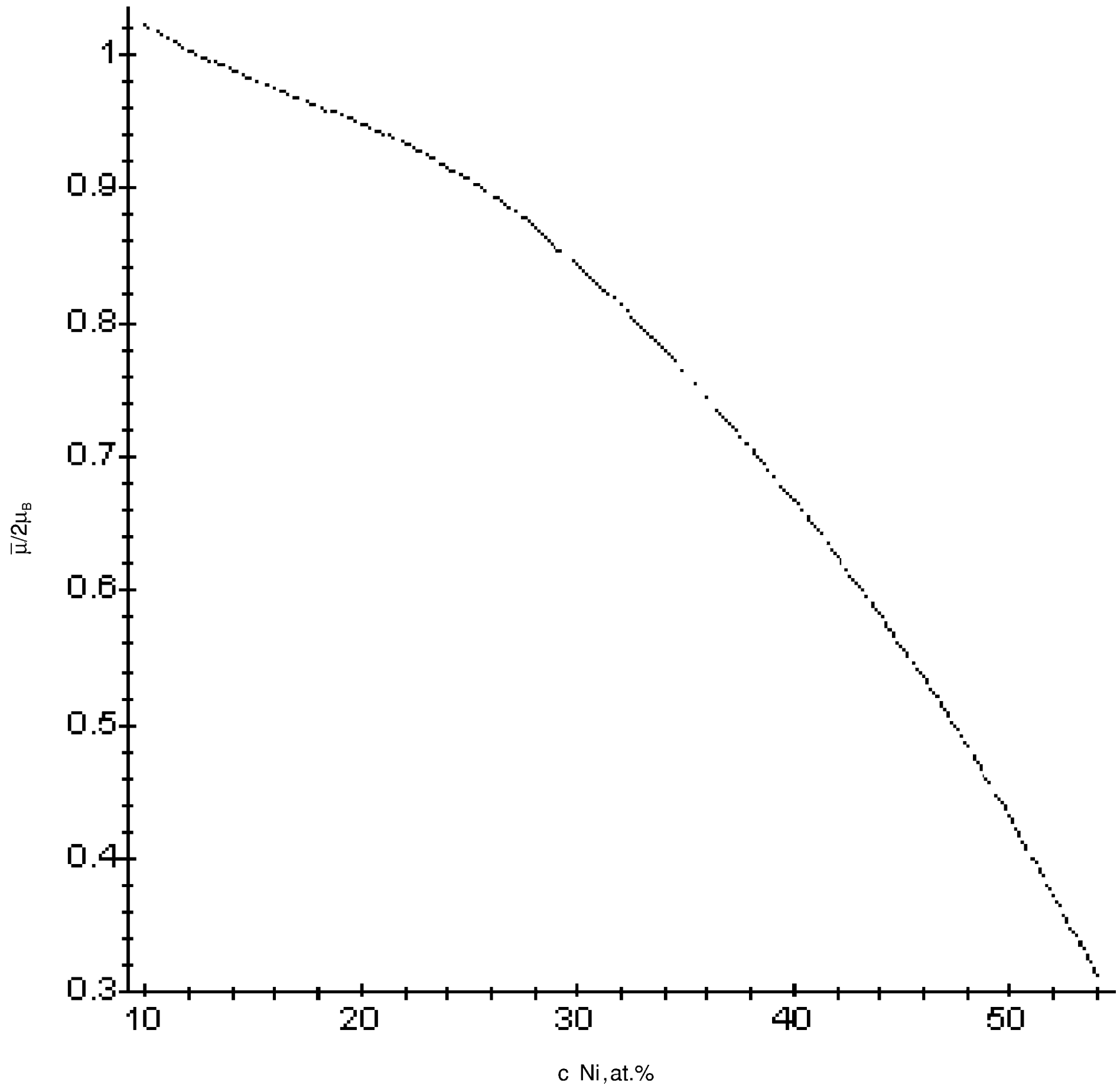

Fig. 5. Concentration dependence of an average 'atomic' magnetic moment, $\overline{\mu}$, of Ni–Fe alloy (with $L1_2$-type superstructure) at $T_K(c_{\rm Fe})$.

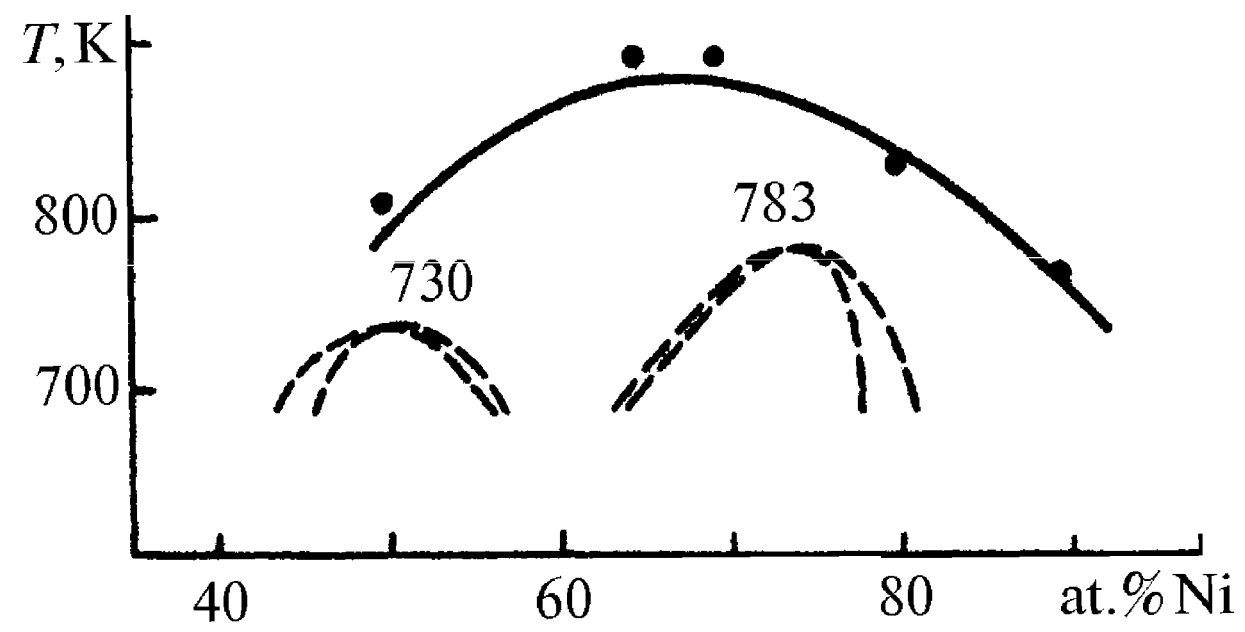

Fig. 6. Numerically calculated by cluster variation method dependencies of Curie (solid line) and Kurnakov (dotted line) temperatures on Ni content for f.c.c.-Ni–Fe alloys [14]; black circles—experimental points [9].